\documentclass[11pt]{article}
   \usepackage{amsmath,amssymb,amsbsy,amsfonts}
   \usepackage[round]{natbib}
   \usepackage[british]{babel}
   \usepackage{fancyhdr}
   \usepackage{geometry}
   \usepackage{graphicx}
\renewcommand{\baselinestretch}{2}

\makeatletter

\def\@makechapterhead#1{\thispagestyle{empty}%
  \vspace*{50\p@}%
  {\parindent \z@ \raggedright \normalfont
    \ifnum \c@secnumdepth >\m@ne
      \if@mainmatter
        \huge\bfseries\centerline{ \@chapapp\space \thechapter}
        \par\nobreak
        \vskip 20\p@
      \fi
    \fi
    \interlinepenalty\@M
    \Huge \bfseries \centerline{#1}\par\nobreak
    \vskip 40\p@
  }}

\@addtoreset{equation}{section}
\@addtoreset{figure}{section}
\@addtoreset{table}{section}

\makeatother

\fancypagestyle{plain}{
			\fancyhead{}
			
}

\newcounter{oldsection}
\renewcommand{\theequation}{\arabic{section}.\arabic{equation}}

\begin{document}

\renewcommand{\baselinestretch}{1}
\title{{\bf Hydrodynamical interaction between an accretion flow and a stellar
       wind}}
\author{S. Mendoza\(^{1,}\)\thanks{Email address: sergio@astroscu.unam.mx},
        J. Cant\'o\(^{1}\) and A.~C. Raga\(^2\)\\
          \(^1\) Instituto de Astronom\'{\i}a, Universidad Nacional 
          Aut\'onoma de M\'exico\\ AP 70-264, Distrito Federal 04510,
          M\'exico\\
          \(^2\) Instituto de Ciencias Nucleares, Universidad Nacional
          Aut\'onoma de M\'exico\\ AP 70-543,  Distrito Federal 04510,
          M\'exico
	  }

\date{\today}

\maketitle

\renewcommand{\baselinestretch}{2}

\begin{abstract}
  Molecular clouds in the interstellar medium suffer gravitational
instabilities that lead to the formation of one or multiple stars.
A recently formed star inside a cold cloud communicates its gravitational
force to the surrounding environment and soon an accretion flow
falling into the star develops.  After their formation, all stars soon
eject a wind of gas that interacts with the external accretion flow.
This interaction produces a shock wave that evolves with time.
The work presented in this article formulates a simple prescription for
the evolution of this interaction.  With the aid of this model we 
construct a few radio continuum maps of the source.  \\ 


\end{abstract}

\section{Introduction}
\label{introduction}

  When a low mass, isolated star is formed inside a gas cloud because of
a gravitational collapse, the process of accreting gas from the cloud onto
the star begins \citep{shu87}.  At a later time a wind of gas emanating
from the star collides with the accretion flow.  Depending on the ram
pressure that the wind generates, it will be able to expand forever (when
the ram pressure of the wind is much greater than the ram pressure of
the medium), reach a steady state (when accretion and wind ram pressures
are comparable) or collapse in the surface of the star (when the accretion
ram pressure is greater than that of the wind) \citep{wilkin98,wilkin03}.

  \citet{bondi52} investigated the form of a spherical accretion flow
induced by the gravitational pull of a central object.  Because of the
spherical symmetry of the accretion quantities, the shocks produced
by the spherical wind and the accretion flow are both spherical and
evolve with time.  \citet{ulrich76} made modifications to the spherical
accretion model by considering a cloud with a small angular momentum.
The cloud is rotating about an axis on which the central star lies.
The result is that the accretion quantities in this model have
cylindrical symmetry, i.e. they do not depend explicitly on the
azimuthal angle.  The interaction between this ``\citeauthor{ulrich76}
flow'' and the spherical wind from the star produces two shocks
that evolve with time.  In this article we develop a simple model for
this particular hydrodynamical interaction.  We do this by equating the ram
pressure of the wind with the ram pressure of the accretion flow.  This is
the easiest way to model the interaction.

  \citet{wilkin03} have already constructed a numerical solution to the
problem by solving the complete hydrodynamical equations.  These authors
have included on their code not only pressure forces that give rise
to the shocks. On their analysis they also included centrifugal forces
that appear because of the curvature of the shock, and they took into
account the gravitational field produced by the central star. The
problem with such a general solution is that it is not very easy to
predict observational quantities.  In the present article we show how,
with a simplified method, it is possible to produce radio continuum maps
of the interaction between the wind and the accretion flow.

   In section \ref{ulrich-model} we describe briefly
\citeauthor{ulrich76}'s accretion model.  Sections
\ref{steady-interaction} and \ref{time-evolution} describe the steady
state and the time evolution of the interaction.  Finally, in section
\ref{astrophysical-consequences} we describe the validity of the predicted
thin layer approximation and we calculate radio continuum maps from
our model.

\section{Ulrich's accretion model}
\label{ulrich-model}

  Let us briefly discuss some of the most important properties of a small
angular momentum accretion flow that falls to a central compact object,
such as a star.  This type of accretion flow was first investigated
by \citet{ulrich76}.

  Consider a star, or any other compact object of mass \( M \).  The star
is embedded on a gas cloud, which for simplicity will be consider to be
of infinite extension.  Far away from the star, the density \( \rho_\infty
\), the pressure \( p_\infty \) and the velocity of sound \( c_\infty \)
have uniform constant values.  We assume that the cloud is rotating
about the \( z \) axis as a solid body in such a way that its specific
angular momentum \( \Gamma \) has the value \( \Gamma_\infty \) far away
from this axis.  We will consider the case in which the accretion process
is so low, that the mass \( M \) of the central object can be taken as
constant.  Finally, we assume that the flow has reached a steady state.
The accretion flow obeys a polytropic relation

\begin{equation}
  \frac{ p }{ p_\infty }= \left( \frac{ \rho }{ \rho_\infty } 
    \right)^\kappa,
\label{eq1.1}
\end{equation}

\noindent where \( p \) and \( \rho \) are the pressure and density of the
gas respectively, and \( \kappa \) is the polytropic index.

  Due to the fact that the accretion rate \( \dot{M} \) in \citet{bondi52}
accretion flow is constant, and under the assumption that the mass of the
central object is much greater than the gas mass contained on a sphere
of radius \( r \) (\( r \) the distance from the star to a certain fluid
particle), the self gravity of the gas is negligible with respect to
the gravity of the central object.  In what follows we will consider \(
\Gamma_\infty \) to be small so that the accretion with rotation can
be considered to be a perturbation from the non-rotational accretion
Bondi flow.

  The problem is characterised by the following parameters: the
gravitational constant \( G \), the radius of the central object \( r_0
\), the constants \( \Gamma_\infty \), \( M \), \( \rho_\infty \), \(
p_\infty \), \( c_\infty \), and the polytropic index  \( \kappa \).  With these
quantities it is possible to build three dimensionless parameters (without
counting \( \kappa \)) 

\begin{equation}
  \delta= \frac{ r_0 c_\infty^2 }{ GM }, \qquad 
    \epsilon = \left( \frac{\Gamma_\infty c_\infty }{GM} \right)^2,
\label{eq.1.2}
\end{equation}

\noindent and \( p_\infty / \rho_\infty c_\infty^2 \sim 1 \).   The length \(
r_\text{c} = GM / c_\infty^2 \)  is (apart from a function that depends of
\( \kappa \) which is of order one) the radial distance where the
fluid particle reaches the velocity of sound in the accretion without
rotation \citep{bondi52}.  In other words, the fact that \( \delta \ll 1
\) ensures that the radius of the central object is small compared with
this distance.

  The dimensionless parameter \( \epsilon \) is ratio of the centrifugal
force to the gravitational force of a fluid particle evaluated at position
\( r_\text{c} \).  Thus, if we impose the condition \( \epsilon \ll 1 \),
the rotational effects are small perturbations to the Bondi accretion
flow.

  Due to the fact that the angular momentum of the accretion flow must
be conserved, eventually the centrifugal force (\( \sim \Gamma^2 /
r^3 \)) will be equal to the gravitational force (\( G M / r^2 \)).
The position \( r_\text{d} = \epsilon r_\text{c} \) where this occurs
is precisely the distance where the flow deviates considerably from the
Bondi spherical accretion and where a disc of radius \( r_\text{d} \)
on the equatorial plane forms \citep{canto95}.

  Low mass, recently formed stars  are such that \( r_0 \sim 2 \,
\mathrm{R}_\odot \), \( c_\infty \sim 1 \, \text{km} \, \text{s}^{-1} \), \( M
\sim 1 \, \textrm{M}_\odot \), with an azimuthal velocity far away from
the star of the order of \( \dot \varphi \sim 10^{-14} \textrm{s}^{-1}
\), and a lifetime \( \tau \) value such that  \( 10^4 \textrm{yrs}
\lesssim \tau \lesssim 10^5 \textrm{yrs} \).  With these values, the
angular momentum far away from the rotation axis takes the value \(
\Gamma_\infty = ( c \tau )^2 \dot{ \varphi } \sim 9.61\times 10^{18} \,
\textrm{cm}^2 \, \textrm{s}^{-1} \).  Thus, for low mass, recently formed
stars, it follows that \(\delta \sim 1.75 \times 10^{-5} \ll 1 \),  \(
\epsilon \sim 0.2 \ll 1 \), \( r_c \sim 536 \, \textrm{AU} \) and \(
r_d \sim 100 \, \textrm{AU} \).

  At time \( \tau = 0 \), when the gravitational effects produced by
the star begin to be noticed by the gas in the cloud, the gas starts to
be accreted.  In what follows we will consider that the time \( \tau
\) is sufficiently large so that \( \int^{r_\text{c}}_0 \mathrm{d}r /
c(r) < \tau \), in which \( c(r) \) represents the local sound speed.
For instance, if the accretion gas is isothermal, the speed of sound
is constant and the previous inequality means that \( r_\text{c} <
c \tau \).  For the general case in which the velocity of sound is not
constant, this argument implies that the points \( r \) contained inside
the sphere determined by \( \int_0^r \mathrm{d} r / c(r) >  \tau \) are
not perturbed by the gravitational effects of the star.  On the other
hand, the gas particle points \( r \) inside the shell that satisfies \(
\int_0^{r_\text{c}} \mathrm{d} r/ c(r) < \int_0^r \mathrm{d} r / c(r)  <
\tau \) are such that they are being accreted to the centre subsonically
and its rotational effects are negligible.  The flow inside the region
\( r_\text{d} < r < r_\text{c} \) is being accreted with a supersonic
velocity and its rotational effects are small.  When \( r < r_\text{d} \),
the rotational effects are very important on the dynamics of the accretion
flow and the velocity of the gas flowing towards the centre is supersonic.

  Because pressure gradients and variations on the internal energy along
the streamlines of the accretion flow inside the sphere of radius \(
r_\text{c} \) contribute little to the energy and momentum equations,
the streamlines can be taken as ballistic.   If we also consider that the
mass of the disc \( M_\text{d} \) is much smaller than the mass of the
central object, then the trajectory of a fluid element of the accretion
flow is described by a central Newtonian potential.

  The total mechanical and internal energy of a fluid element far away
from the central object is non--zero.  This is because the azimuthal
component of the velocity and the temperature differ from zero at that
point.  Nevertheless, the difference between the total energy and the
kinetic plus potential energy of a fluid element near the disc is small.
This is valid for small values of the rotational kinetic energy (i.e. 
when \( \epsilon \ll 1 \)) and when heating due to radiation is
insignificant.  In other words, the streamlines near the disc have to
be parabolic trajectories such that

\begin{equation}
  E = \frac{ 1 }{ 2} v^2 - \frac{ GM }{ r } \approx 0,
\label{eq.1.3}
\end{equation}

\noindent where \( E \) is the energy per unit mass of a fluid element
near the disc and \( v \) its velocity.  Once the particles reach the
plane \( \theta = \pi / 2 \) (\( \theta \) the polar angle) where the
disc lies, the particles merge into the disc.  Because of this, the
only interesting trajectories for the accretion flow are the upstream
trajectories to the disc.

  For simplicity, in what follows we will work with dimensionless
quantities, unless stated otherwise.  In other words, let us make the
changes

\begin{equation}
   \frac{ r }{ r_d} \longrightarrow r \ ,\qquad \frac{ v_i }{ v_\text{k} }
      \longrightarrow  v_i,\quad \left(i = r \ , \theta, \varphi
      \right) \qquad \frac{\rho }{ \rho_0 } \longrightarrow \rho,
\label{eq.1.10}
\end{equation}

\noindent where the natural velocity \( v_\text{k} \)\footnote{ \(
v_\text{k} \) is the velocity (keplerian velocity) of a particle that
rotates about an axis on a circular trajectory of radius \( r_\text{d}
\) under the influence of a central Newtonian field. } and density \(
\rho_0 \) are defined by

\begin{equation}
  v_\text{k}  = \left( \frac{ GM }{ r_d } \right) ^ { 1 / 2},\qquad 
    \rho_0 = \frac{ \dot{M} }{ 4 \pi r_d ^2 v_\text{k} }.
\label{eq.1.12}
\end{equation}

\begin{figure}
  \begin{center} 
    \includegraphics[scale=0.7]{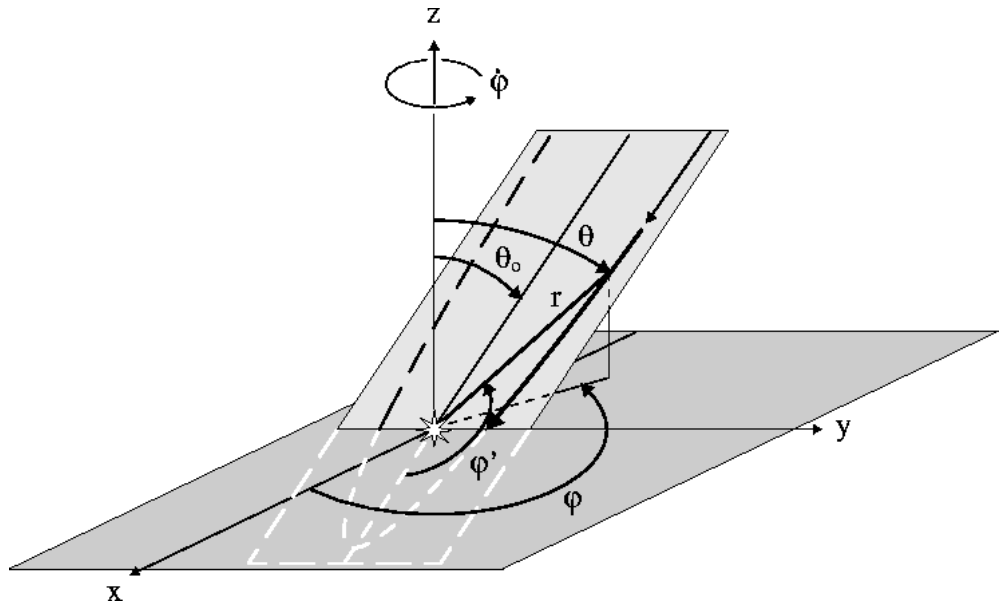}
  \end{center}
  \caption[Accretion Scheme]{ Far away from the star, the fluid particles
	   in Ulrich's accretion rotate like a solid body with constant
	   angular velocity \( \dot \varphi \). These particles follow
	   parabolic trajectories near the central object (star).  Only
	   upstream trajectories to the equatorial plane are of interest.}
\label{fig.1.1}
\end{figure}

\noindent The analytical form of the stream lines, the velocity field
and density for the accretion flow with rotation are given by 
\citep{ulrich76}

\begin{gather}
  r = \frac{ \sin^2 \theta_0 }{  1 - \cos\theta / \cos\theta_0 },
							\label{eq.1.13} \\
  v_r = - \left( \frac{ 1 }{ r } \right) ^{ 1 / 2 } \left( 1 + \frac{
    \cos \theta }{ \cos \theta_0 } \right)^{ 1 / 2 },
    							\label{eq.1.14} \\
  v_\theta = \left( \frac{ 1 }{ r} \right)^{ 1 / 2 }  \frac{ \cos \theta_0
    - \cos \theta }{ \sin \theta } \left( 1 + \frac{ \cos \theta }{
    \cos\theta_0 } \right) ^{ 1 / 2 },
    							\label{eq.1.15} \\
  v_\varphi = \left( \frac{ 1 }{ r } \right)^{ 1 / 2 } \left( 1 - \frac{
    \cos \theta }{ \cos \theta_0 } \right)^{ 1 / 2} \frac{ \sin \theta_0
    }{ \sin \theta },
    							\label{eq.1.16} \\
  \rho = r  ^{- 3 /  2} \left( 1 + \frac{ \cos \theta }{ \cos \theta_0 }
    \right)^{ -1 / 2} \left\{ 1 + 2 r ^{ -1 } P_2 \left( \cos \theta_0
    \right)  \right\}^{-1},
    							\label{eq.1.17} 
\end{gather}

\noindent where \( v_i \) (\(i=r,\theta,\varphi \)),  and \( \rho \)
are the velocity and density fields of the accretion flow respectively.
The azimuthal angle is represented by \( \varphi \).  The initial polar
angle \( \theta_0 \) that a fluid element has once it begins its way
down towards the disc (see Figure~\ref{fig.1.1}), \( \dot M \) is the
accretion rate and \( P_2 \left( \xi \right) \) is the second order
Legendre polynomial given by

\begin{equation}
  P_2\left( \xi \right) = \frac{ 1 }{ 2 } \left( 3 \xi ^2 - 1 \right).
\label{eq.1.9}
\end{equation}

\begin{figure}
  \begin{center} 
    \includegraphics[scale=0.6]{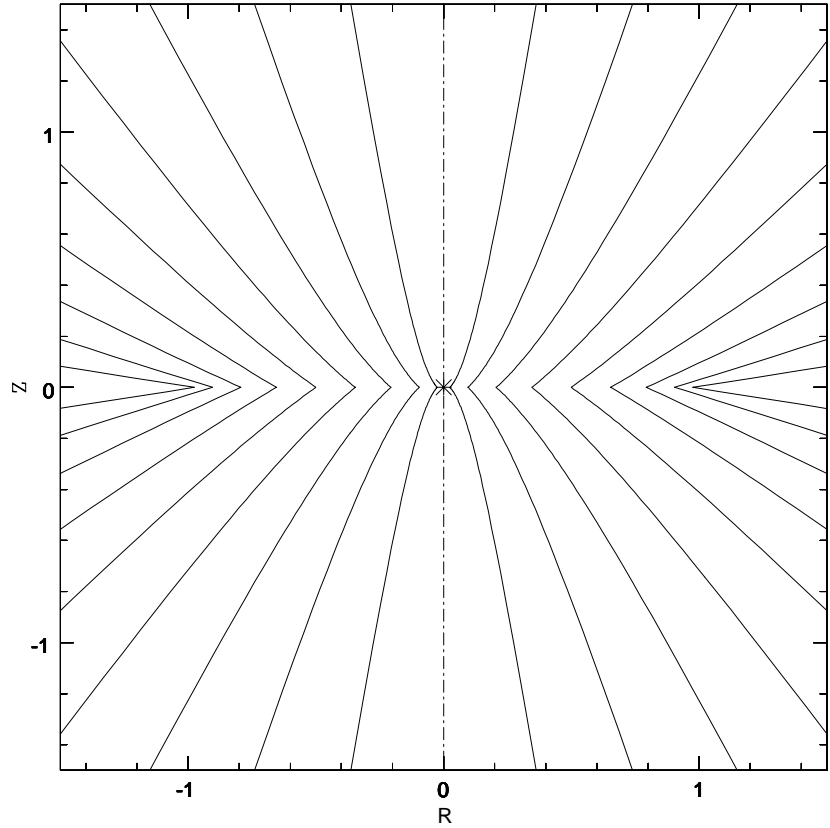}
  \end{center}
  \caption[Streamlines for accretion flow]{ The figure shows the
	   streamlines of the accretion flow with rotation projected on a
	   plane with constant azimuthal angle.  Every single streamline
	   is labelled by its initial polar angle \( \theta_0 \), which is
	   the initial particle's angle with respect to the \( Z \) axis as
	   the gas begins to be accreted to the star.  Lengths in the
	   diagram are measured in units of the radius of the disc \(
	   r_\text{d} \). }
\label{fig.1.2}
\end{figure}

  A plot of the streamlines (Eq.~\eqref{eq.1.13}) projected on a plane \(
\varphi = \text{const} \) is shown in Figure~\ref{fig.1.2}.  The angle \(
\theta_0 \) labels a particular streamline.  For each \( \theta_0 \) there
is only one streamline, because they cannot intersect.  It is clear from
the velocity field equations that when \( \theta_0 \rightarrow 0 \),
the polar and azimuthal components of the velocity become zero as well.
Thus, the streamlines become parallel near the axis of rotation.


  The velocity and density field equations are functions that depend
only on the particle's position and not of the initial polar angle \(
\theta_0 \).  This is easily seen by rewriting Eq.~\eqref{eq.1.3} as

\begin{equation}
   \cos^3 \theta _0 + \cos\theta _0 (r - 1) -r \cos \theta = 0,
\label{eq.1.18}
\end{equation}

\noindent which has a solution (see appendix \ref{appendix})

\begin{equation}
    \cos \theta_0 =
    \begin{cases}
        \left( \cos \theta \right)^{ 1 / 3 },\\
			\qquad	\qquad \thickspace
				\text{for } r = 1,    \\
        2 \left( \frac{ r - 1 }{ 3 } \right)^{ 1 / 2 } \sinh \left\{ \frac{
	  1 }{ 3 } \textrm{arcsinh} \left( \frac{ r \cos \theta }{ 2 
	  \left( \frac{
	  r - 1 }{ 3 } \right)^{ 3 / 2 } } \right) \right\}, \\
			\qquad	\qquad \thickspace
				\text{for } r > 1,    \\
    2 \left( \frac{1 - r}{ 3 } \right) ^ {1 / 2} \cosh \left\{ \frac{
      1 }{ 3 } \textrm{arccosh} \left( \frac{ r \cos \theta }{ 2 \left(
      \frac{ 1 - r }{ 3 } \right)^{3 / 2} } \right) \right\}, \\ 
			\qquad	\qquad \thickspace
			\text{for }
				r < 1 \text{ and } \left( \frac{
				r \cos \theta }{ 2 } \right)^2 -
				\left( \frac{ 1 - r }{ 3 } \right)^3 >
				0,        \\
    2 \left( \frac{ 1 - r }{ 3 } \right) ^ { 1 / 2 } \cos \left\{ \frac{
      1 }{ 3 } \textrm{arccos} \left(  \frac{ r \cos \theta }{ 2 \left( \frac{
      1 - r }{ 3 } \right)^{ 3 / 2 } } \right) \right\}, \\
			\qquad	\qquad \thickspace
			\text{for }
				r < 1 \text{ and }  \left( \frac{ r \
				\cos \theta }{ 2 } \right)^2 - \left(
				\frac{ 1 - r }{ 3 } \right)^3 < 0.
\end{cases} 
\label{eq.1.19}
\end{equation}

  Thus, Eqs.~\eqref{eq.1.14}-\eqref{eq.1.17} and Eq.~\eqref{eq.1.19} provide
the velocity and density fields as functions of the polar angle \( \theta
\) and the radial coordinate \( r \) only.

  From Eq.~\eqref{eq.1.18} it is easy to see that

\begin{gather}
  \cos \theta_0 \bigg|_{ \theta = \frac{ \pi }{ 2 } } = \Theta \left(
    1 - r \right) \sqrt{ 1 - r },
						\label{eq.1.20} \\
  \cos{ \theta_0 } =  1 - \frac{ \theta^2 }{ 2 } \left( \frac{ r 
    }{ r + 2 } \right), \quad \text{ as } \theta \rightarrow 0,
    						\label{eq.1.21}
\end{gather}

\noindent where \( \Theta(\chi) \) is the Heaviside step function with
a value of \( 1 \) for \( \chi > 0 \) and \( 0 \) for \( \chi < 0 \).
With the substitution of Eq.~\eqref{eq.1.20} and Eq.~\eqref{eq.1.21} into
Eq.~\eqref{eq.1.17} it is possible to give values for the density in the
equatorial plane and the polar axis respectively

\begin{gather}
  \rho \left( \theta = \pi / 2 \right) =
     \begin{cases} 
       \frac{ 1 }{ 2 } r^{-1/2} \left( 1 - r \right)^{-1} ,&   
       				\text{for } r<1,   \\
       \left( 2 r - 1 \right)^{ - 1 / 2 } 
         \left( r - 1 \right)^{-1} ,&   
	 			\text{for } r \geq 1, 
     \end{cases}
     						\label{eq.1.22}   \\
  \rho \left( \theta = 0 \right) = \left(  2 r  \right)^{-1/2} \left( r + 2 
    \right)^{-1} .
  						\label{eq.1.23}
\end{gather} 

  The fact that the density tends to infinity as \( r \) tends to zero for
any \( \theta \) is due to the accumulation of the accreted gas around
the central object.  Nevertheless, the border of the disc (i.e. \( r =
1 \) and \( \theta = \pi/2 \)) has also an infinite density.  This is
because we are considering an infinitely thin disc and so, border effects
on it are expected to appear.

\begin{figure}
  \begin{center} 
    \includegraphics[scale=0.6]{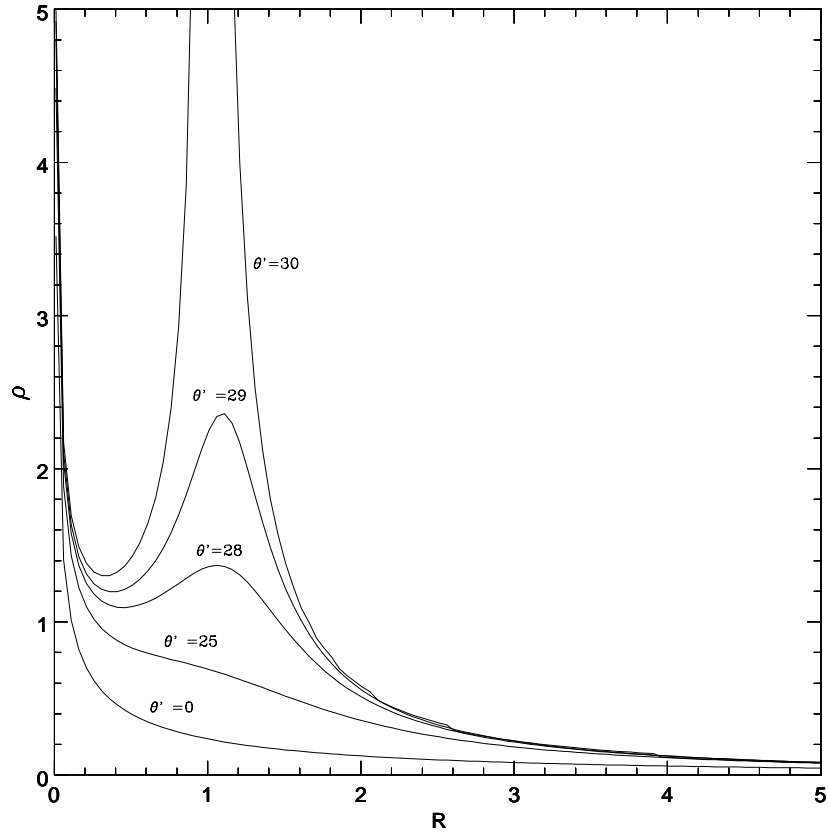}
  \end{center}
  \caption[Density of the accretion flow]{ Accretion density \( \rho \)
	    varies continuously from \( \theta = 0 \) to \( \theta = \pi /
	    2 \). The density tends to infinity in the point  \( r = 1 \)
	    for the \( \theta = \pi / 2 \) curve which corresponds to
	    the border of the disc.  \( R \) represents the cylindrical
	    radius (\( R \equiv r \sin \theta \)). Lengths are measured
	    in units of the radius of the disc (\( r_d \)) and the
	    density is measured in units of \( \rho_0 \) (see text).
	    The polar angle \( \theta' \) is such that \( \theta =
	    (\pi / 60)\theta'\). }
\label{fig.1.3}
\end{figure}

\begin{figure}
  \begin{center} 
    \includegraphics[scale=0.6]{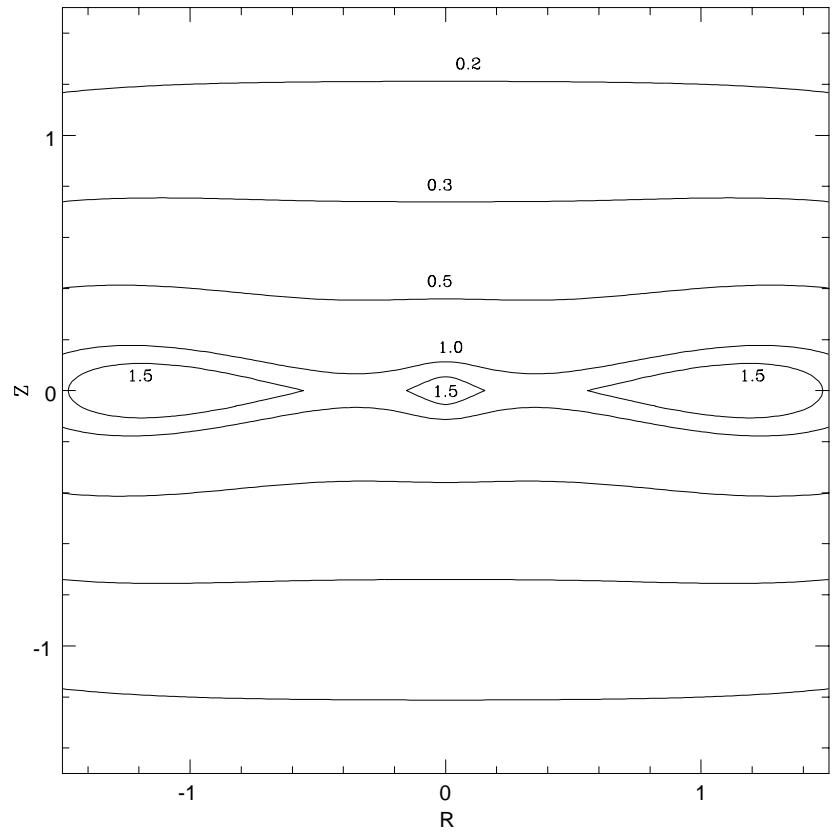}
  \end{center}
  \caption[Density isocontours of the accretion flow]{ Density isocontours
	   are shown as continuous lines for the accretion flow.
	   The density \( \rho \) is measured in units of the density \(
	   \rho_0 \).  Lengths are measured in units of the radius of
	   the disc. Different numbers in the plot represent different
	   values of the density.}
\label{fig.1.4}
\end{figure}

  Figure~\ref{fig.1.3} shows the variation of the density as a function
of position for different polar angles.  This plot, together with the
density contours (as shown by Figure~\ref{fig.1.4}) can be used to make
a detailed analysis of the density distribution in the accretion flow.

\section{Steady interaction}
\label{steady-interaction}

  From now on, let us suppose that the central object is a recently formed
star, and that a supersonic, spherical wind with constant radial velocity
is being produced by it.  In this section we discuss the interaction
between the Ulrich accretion flow described in Section \ref{ulrich-model}
with the wind of the star.

  The interaction between the supersonic accretion flow and the supersonic
stellar wind gives rise to an initial discontinuity.  The end result of
this interaction is the production of two shock waves and a tangential
discontinuity between both.  In the frame of reference of the tangential
discontinuity, the shocks separate from each other.  Under the assumption
that both shocks occupy the same position in space, so that we can call
both shocks ``the shock'' for simplicity, we analyse the geometrical form
that the shock has under the assumption that the only type of forces that
keep the shock stable are pressure forces.  Apart from pressure forces,
there will be centrifugal forces (produced by the post--shock material)
acting on the walls of the shocks.  These forces are originated because
the shock is curved.  In section \ref{astrophysical-consequences} we
analyse under which conditions, the assumption that both shocks occupy
the same position in space (thin layer approximation) is valid by studying
the cooling lengths of the flow.  It turns out that for the steady case,
this approximation is correct.

  Under these assumptions, the locus of the points of the shock are given
by balancing the ram pressure produced by both flows, the accretion flow
and the stellar wind.  In other words, the shock is located at points for
which 

\begin{equation}
  \rho_\text{w} v_\text{wn}^2 = \rho v_\text{n}^2,
\label{eq.2.1}
\end{equation}

\noindent where the subindex \( \text{w} \) labels quantities related to
the stellar wind and \( n \) means the normal component of the velocity.
The wind's density is determined by the mass loss rate \( \dot{M}_\text{w}
\) of the star

\begin{equation}
  \dot{M}_\text{w} = 4 \pi r^2 v_\text{w} \rho_\text{w}.
\label{eq.2.2}
\end{equation}

  In order to get a dimensionless form of Eq.~\eqref{eq.2.1}, we make the
changes

\begin{equation}
  \frac{ v_\text{wn} }{ v_\text{w} } \longrightarrow v_\text{wn}, \qquad
    \frac{ \rho_\text{w} }{ \rho_{ \text{w} 0 } } \longrightarrow
    \rho_\text{w},
\label{eq.2.3}
\end{equation}

\noindent where \( \rho_\text{w0} = \dot{M}_\text{w} / 4 \pi r_\text{d}^2
v_\text{w} \). Combining these changes with the ones presented in 
Eq.~\eqref{eq.1.10} we obtain

\begin{equation}
  \rho_\text{w} v^2_\text{wn} \lambda = \rho v^2_\text{n}
\label{eq.2.4}
\end{equation}

\noindent where \( \lambda \) is a dimensionless parameter given by

\begin{equation}
  \lambda \equiv \frac{\dot{ M }_\text{w} v_\text{w} }{ \dot{ M }
    v_\text{k} }.
\label{eq.2.5}
\end{equation}

  Eq.\eqref{eq.2.4} determines the position \(
\boldsymbol{r}(\theta,\varphi) \) of the shock as a function of the polar
angle \( \theta \) and the azimuthal angle \( \varphi \).   A normal
vector to this surface is given by

\begin{equation}
  \boldsymbol{n} = \frac{ \partial \boldsymbol{r} }{ \partial \theta } 
    \times \frac{ \partial \boldsymbol{r} }{\partial \varphi},
\label{eq.2.6}
\end{equation}

\noindent since \( \partial \boldsymbol{r} / \partial \theta \) and \(
\partial \boldsymbol{r} / \partial \varphi \) are tangent vectors to
any surface parametrised by the polar and azimuthal angles.  Because the
accretion flow and the stellar wind have azimuthal symmetry, the shock
wave position should not depend on the angle \( \varphi \).  Since the
length element \( \mathrm{d} \boldsymbol{r} = \mathrm{d}r \boldsymbol{e}_r
+ r \mathrm{d}\theta \boldsymbol{e}_\theta + r \sin \theta \mathrm{d}
\varphi \boldsymbol{e}_\varphi \), where \( \boldsymbol{e}_i \) (\(i
= r,\theta,\varphi\)) are unitary vectors in the i--th direction, then
Eq.~\eqref{eq.2.6} can be rewritten in the following form

\begin{equation}
  \widehat{\boldsymbol{n}} = \frac{ r \boldsymbol{e}_r - \left( \partial
    r / \partial \theta \right) \boldsymbol{e}_\theta }{ { \left\{  r^2 +
    \left( \partial r / \partial \theta \right)^2 \right\} }^{ 1 / 2 } },
\label{eq.2.7}
\end{equation}

\noindent where \( \widehat{\boldsymbol{n}} \equiv \boldsymbol{n} / n \) is
a unit vector in the direction of \( \boldsymbol{n} \).

  The accretion flow and wind velocities are radial on the rotation
axis. This means that the locus of the shock satisfies the following
boundary condition

\begin{equation}
  \frac{ \partial r }{ \partial \theta } \bigg|_{ \theta = 0 } = 0.
\label{eq.2.8}
\end{equation}

\noindent Using Eq.~\eqref{eq.2.7} and Eq.~\eqref{eq.2.8}, it follows from
Eq.~\eqref{eq.2.4} that over the rotation axis of the cloud, 
the value of the distance \( r \) from the star to the shock is given by

\begin{equation}
  r \bigg|_{ \theta = 0 } = -2 + \frac{ 1 - \sqrt{ 1 -
    4 \lambda^2 } }{ \lambda^2 }.
\label{eq.2.9}
\end{equation}

\noindent This equation implies that for \( \lambda > 1/2 \) there is no
value for the initial condition of the position of the shock. In other
words, there is no steady solution for the case when \( \lambda > 1/2 \).

  The equation satisfied by the spatial points on the shock \( r(\theta)
\) is given by substitution of  Eq.~\eqref{eq.2.6} into Eq.~\eqref{eq.2.4},
that is

\begin{equation}
  \frac{ \partial r }{ \partial \theta } = \frac{ \lambda^{ 1 / 2 } +
    \rho^{ 1 / 2 } r v_r }{ \rho^{ 1 / 2 } v_\theta }. 
\label{eq.2.10}
\end{equation}

\begin{figure}
  \begin{center}
    \includegraphics[scale=0.6]{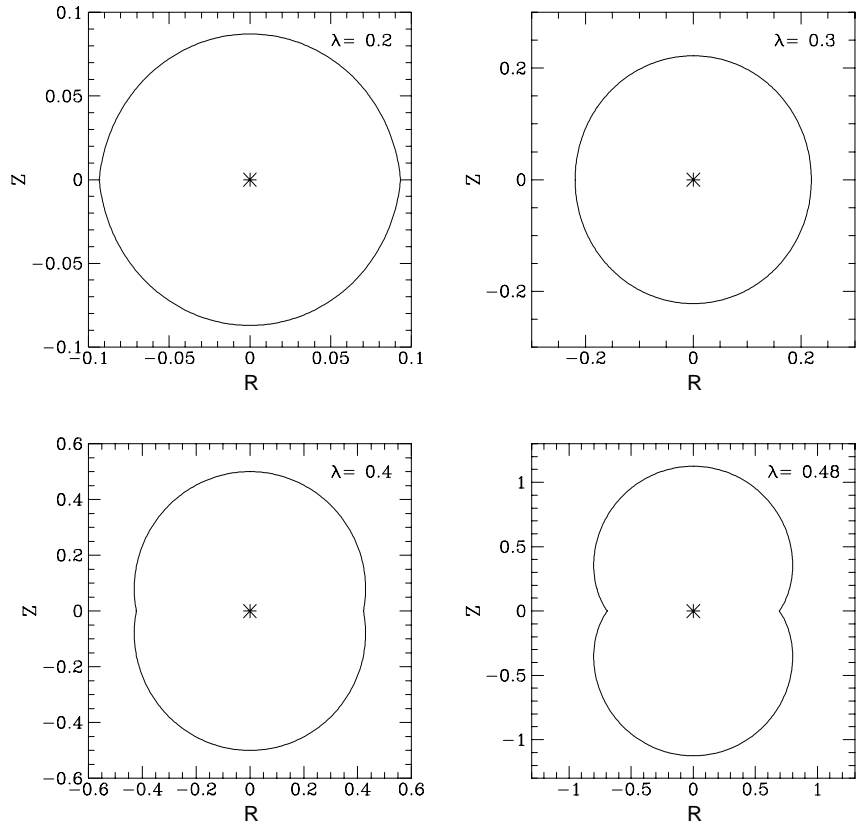}
  \end{center}
  \caption[Geometrical shape of the steady shock]{ The interaction between
           the stellar wind and the accretion flow produces two strong
	   shock waves.  It is assumed that both shocks occupy the same
	   position in space.  \( R \) represents the cylindrical radius
	   (\( R \equiv r \sin \theta \)), and \( Z \) is the polar axis
	   coordinate.  Distances are measured in units of the radius 
	   of the disc \( r_\text{d} \). }
\label{fig.2.1}
\end{figure}

\noindent Figure~\ref{fig.2.1} shows integrals of Eq.~\eqref{eq.2.10}
for different values of the parameter \( \lambda \).  These integrals
were obtained with a 4th rank Runge--Kutta method.  From the figure it
follows that when \( \lambda \leq 1/2 \) the shocks can be classified in
two cases.  Those that have positive derivatives with respect to the polar 
angle (\( \lambda = 0.2 \), for example) and those for which this
derivative is negative (\( \lambda = 0.48 \), for example).  Let us
show that there is no configuration for which this derivative is zero
for all polar angles.  For this, it is sufficient to assume that the
configuration is a circumference.  It then follows that \( r( \theta = 0 )
= r( \theta = \pi / 2 ) \).  From this and Eq.~\eqref{eq.2.9} it follows
that the obtained value for \( \lambda \) does not correspond to a circle.

\begin{figure}
  \begin{center}
    \includegraphics[scale=0.6]{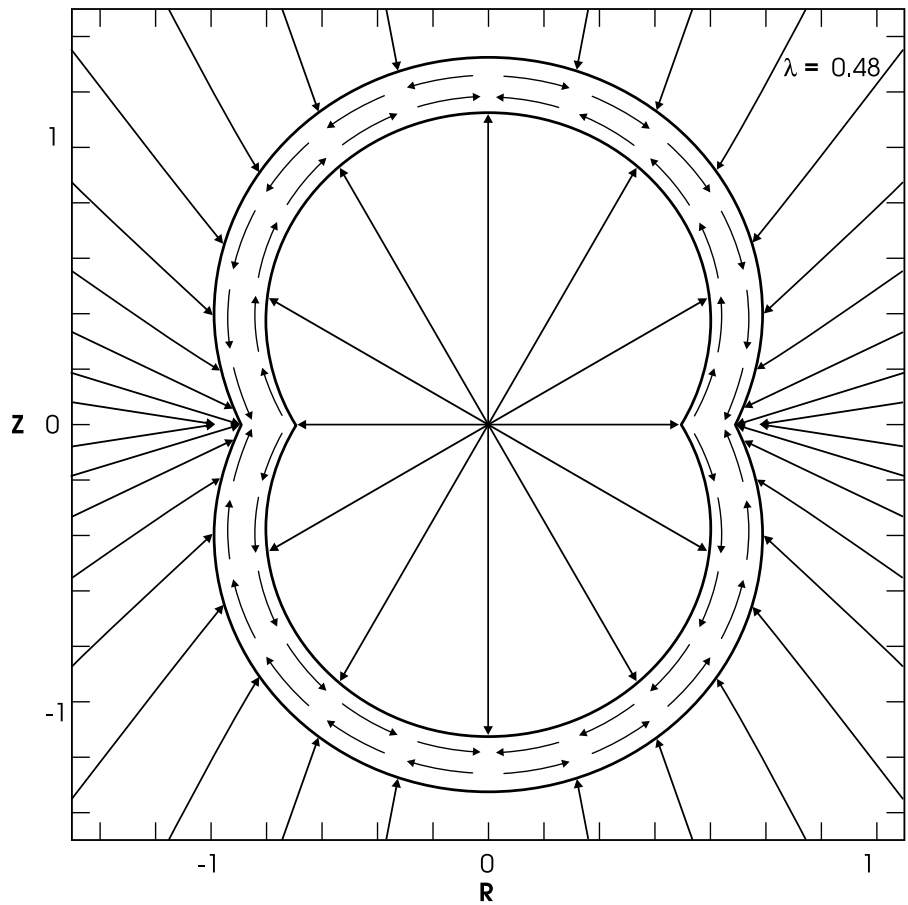}
  \end{center}
  \caption[Wind and accretion shocks]{ The interaction between the
	   stellar wind and the accretion flow forms two shock cavities.
	   The figure shows a schematic representation of this interaction
	   with both shocks lying close to each other.	The arrows in the
	   figure show the direction of the flow velocity immediately
	   after crossing the shocks.  The cylindrical radius \( R \)
	   and the polar coordinate \( Z \) are both measured in units
	   of the accretion disc radius \( r_\text{d} \).}
\label{fig.2.2}
\end{figure}

  Until now, we have considered that both shocks, the accretion shock and
the wind shock, occupy the same position in space.  However, as shown by the
cartoon of Figure~\ref{fig.2.2}, there is an intermediate region of shocked
gas between them.  Let us calculate the direction that the post--shocks
should have immediately after crossing their corresponding shocks.  Under
the assumption that both shocks are highly radiative, the post--shock
normal velocities are negligible.  Nevertheless, since the tangential
components of the velocity are continuous through the shock, the
post--shock direction is determined by the tangential components of the
pre--shock velocity.  Because the shock is not a function of the azimuthal
angle, a unitary tangent vector to it's surface is given by

\begin{equation}
  \widehat{\boldsymbol{\tau}} = \frac{ \left( \partial r / \partial \theta
    \right) \boldsymbol{e}_r + r \boldsymbol{e}_\theta }{ { \left\{  r^2 +
    \left( \partial r / \partial \theta \right)^2 \right\} }^{1 / 2} }.
\label{eq.2.11}
\end{equation}

\begin{figure}
  \begin{center}
    \includegraphics[scale=0.6]{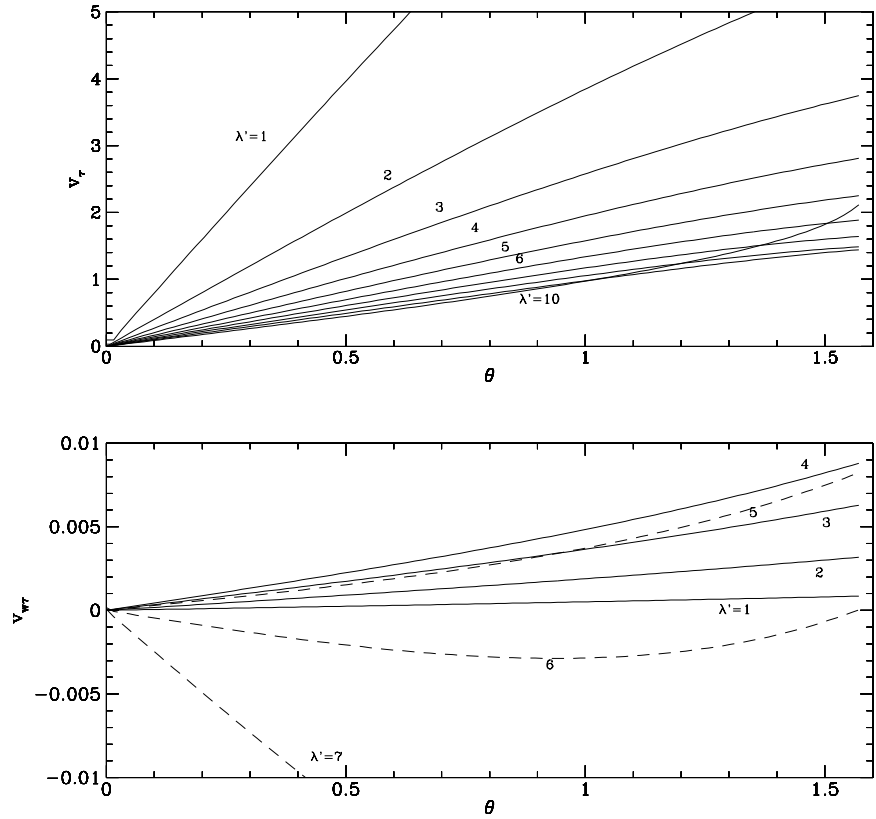}
  \end{center}
  \caption[Post--shock flow direction]{ The post--shock direction is
           given by the values of the pre--shock tangential velocity
	   values for each flow.  The tangential velocity of the stellar
	   wind \( v_{ \text{w} \tau } \) is measured in units of the
	   wind's velocity  \( v_\text{w} \).  The accretion velocity \(
	   v_\tau \) is measured in units of the velocity \( v_\text{k} \)
	   The numbers on the plot are values of the quantity \( \lambda'
	   \) defined by \( \lambda = 0.05 \lambda' \). }
\label{fig.2.3}
\end{figure}

  Due to the symmetry of the problem, it is useful to analyse only the
region \( 0 \leq \theta \leq \pi / 2 \).  Using the tangent vector of
Eq.~\eqref{eq.2.11} and because the polar component of the velocity is
positive for the accretion flow, we say that the post--shock flow (either
accretion or stellar wind) ``goes up'' if \( \boldsymbol{v} \cdot
\hat{\boldsymbol{\tau}} < 0 \) (i.e., the flow leaves the equatorial plane)
and the flow ``goes down'' if \( \boldsymbol{v} \cdot
\hat{\boldsymbol{\tau}} > 0 \) (i.e., the flow approaches the equatorial
plane).  Figure~\ref{fig.2.3} shows values of the pre--shock tangential
velocities for both flows as a function of the polar angle for different
values of the parameter \( \lambda \).  From this figure it follows that in
the case of the stellar wind, for \( \lambda \lesssim 0.3 \) the post--shock
flow goes down and for \( \lambda \gtrsim 0.3 \) this post--shock flow goes
up.  When \( \lambda \approx 0.3 \) the post--shock flow goes up in some
regions and goes down in some others.  For the accretion flow, the
situation is different.  It always goes down, regardless of the
configuration chosen.  It is important to note that there is an
accumulation point of material, corresponding to the polar axis (\( \theta
= 0 \)).  In this point,  both post--shock flows have no motion.  In this
accumulation point, it is expected that the post--shock flow collimates
to regions away from the star when the centrifugal forces are taken into
account.


\section{Time evolution}
\label{time-evolution}

  Let us now analyse the evolution in time of the interaction between
the stellar wind  and the accretion flow.  As discussed in section
\ref{steady-interaction}, two shocks are formed.  For simplicity we
will consider that both shocks occupy the same position in space and
we will refer to this region as the shock, unless stated otherwise.
If we assumme that only ram pressure forces act on the surface of the
shock, then from the system of reference of the shock Eq.~\eqref{eq.2.1}
is also valid.  In order to transform it to the frame of reference of
the star, we make the transformation

\begin{equation}
  v_\text{wn} \longrightarrow \left( v_\text{wn} - v_\text{sn}
    \right), \qquad v_\text{n} \longrightarrow \left( v_\text{sn} -
    v_\text{n} \right),
\label{eq.3.1}
\end{equation}

\noindent where \( v_\text{wn},\ v_\text{sn},\ v_\text{n} \) represent
the velocities of the stellar wind, the shock wave and the accretion
flow respectively, all normal to the shock wave.  With these changes,
Eq.~\eqref{eq.2.1} gives

\begin{equation}
  \rho_\text{w} \left( v_\text{wn} - v_\text{sn} \right)^2 = \rho \left(
    v_\text{sn} - v_\text{n} \right)^2.
\label{eq.3.2}
\end{equation}

  In order to write Eq.~\eqref{eq.3.2} in a dimensionless form, we
use the changes described in Eq.~\eqref{eq.1.10} and Eq.~\eqref{eq.2.3}
together with

\begin{equation}
  \frac{ v_\text{sn} }{ v_\text{k} } \longrightarrow v_\text{sn}.
\label{eq.3.3}
\end{equation}

\noindent Performing these transformations, Eq.~\eqref{eq.3.2} can be
written as

\begin{equation}
  v_\text{sn} = \frac{ \left( \lambda \rho_\text{w} \right)^{ 1 / 2 }
    v_\text{wn} + v_\text{n} \rho^{ 1 / 2 } }{ \left( \eta \rho_\text{w}
    \right)^{ 1 / 2 } + \rho^{ 1 / 2 } },
\label{eq.3.4}
\end{equation}

\noindent where \( \lambda \) is defined by Eq.~\eqref{eq.2.5}
and \( \eta \) is another dimensionless parameter given by

\begin{equation}
  \eta \equiv {\lambda} { \left( \frac{ v_\text{k} }{ v_\text{w} }
    \right)^2}.
\label{eq.3.5}
\end{equation}

\noindent When \( v_\text{sn} \rightarrow 0 \), that is when the
evolutionary shock becomes steady, Eq.~\eqref{eq.3.4} converges to
Eq.~\eqref{eq.2.4}.  

  Since we are only interested in the geometrical shape of the shock and
not on the particular trajectory that a small fragment of the shock
follows, we analyse only the form that the shock will have as a function of
time for a fixed polar angle.  Because of this, the relation between the
shock's radial velocity \( v_\text{s} \) and its normal velocity can be
obtained from Eq.~\eqref{eq.2.7}

\begin{equation}
  v_\text{sn} = \frac{ r v_\text{s} }{ \left\{ r^2 + \left( 
    \partial r / \partial \theta  \right)^2 \right\}^{ 1 / 2 } }.
\label{eq.3.6}
\end{equation}

\noindent With this, Eq.~\eqref{eq.3.4} can be rewritten as

\begin{equation}
  v_\text{s}= \frac{ \partial r }{ \partial t } = \frac{ \left(
    \lambda \rho_\text{w} \right)^{ 1 / 2 } r + \rho^{ 1 / 2 } \left\{
    r v_\text{r} - v_\theta \partial r / \partial \theta \right\} }{
    r \left\{ \left( \eta \rho_\text{w} \right)^{ 1 / 2 } + \rho^{ 1 /
    2 } \right\} },
\label{eq.3.7}
\end{equation}

\noindent where the normal values of the stellar wind and the accretion
flow given by Eq.~\eqref{eq.2.7} have been used.

  Near the star, for \( r \ll 1 \), the velocity field and density for the
accretion flow are spherically symmetric and coincide with the spherical
symmetry of these quantities for the stellar wind.  In other words, at time
\( t = 0 \) the shock is a spherical surface of radius \( r_\star \), the
radius of the star

\begin{equation}
  r \left( \theta, t=0 \right) = \frac{ r_0 }{ r_\text{d} } \ll 1.
\label{eq.3.8}
\end{equation}

  Since Eq.~\eqref{eq.3.4} has two dimensionless parameters, we will
consider in what follows that the star is a recently formed star of low
mass.  For these stars \citep{black85} \( v_\text{w} \sim 300 \,
\textrm{km} \, \textrm{s}^{-1}  \), \( M \sim 1 M\odot \) y \( r_\text{d}
\sim 100 \, \textrm{au} \).  With these values, the parameter  \( \eta \)
can be determined uniquely by the values of \( \lambda \) through the
following expression

\begin{equation}
  \eta = 10^{-4}  \lambda.
\label{eq.3.9}
\end{equation}

\begin{figure}
  \begin{center}
    \includegraphics[scale=0.6]{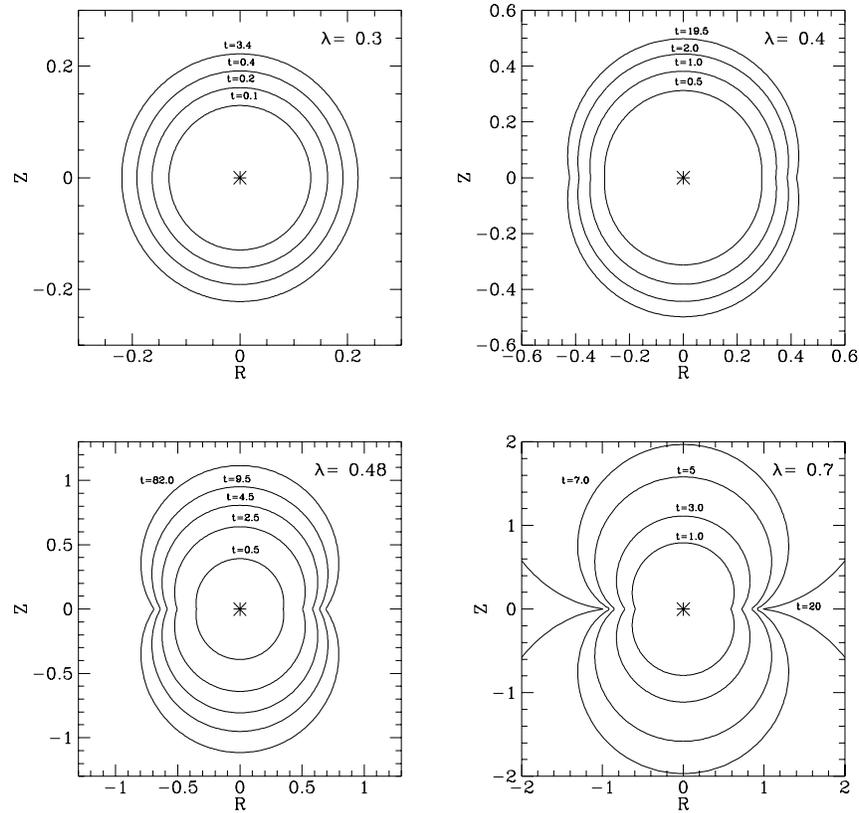}
  \end{center}
  \caption[Time evolution of the shock]{ The figure shows the shape of
	   the shock for different time intervals \( t \), in units
	   of  \( r_\text{d} / v_\text{k} \) and different values of
	   \( \lambda \).  The shock converges to the steady case of
	   Figure~\ref{fig.2.1} when \( \lambda \leq 1/2 \) and for
	   sufficiently large values of time \( t \).  When \( \lambda
	   > 1/2 \) there is no steady configuration and the shock
	   grows indefinitely.	The cylindrical radius \( R \) and the
	   coordinate \( Z \) are measured in units of the radius of
	   the disc \( r_\text{d} \). }
\label{fig.3.1}
\end{figure}

  The solution of Eq.~\eqref{eq.3.7} with the initial condition given
by Eq.~\eqref{eq.3.8} was obtained by numerical integration using
\citet{macormack72} corrector method.  The integrals for this equation are
plotted in Figure~\ref{fig.3.1}.  The solutions show that the steady state
is only reached for \( \lambda \leq 1/2 \), and so there is a convergence
of these solutions to the ones obtained in the previous section.
When \( \lambda > 1/2 \) the shock grows indefinitely, apart from the
equatorial plane, where the shock reaches the border of the disc and
stays there for sufficiently large times (see the case \( \lambda = 0.7 \)
of Figure~\ref{fig.3.1} for example).  This effect is due to the infinite
value that the accretion density reaches in the border of the disc.

  Using Eqs.~\eqref{eq.1.13}-\eqref{eq.1.17} for \( r \gg 1 \) and
Eq.~\eqref{eq.3.7} it is possible to obtain the relation that the shock
surface must have for sufficiently large times

\begin{equation}
  { \frac{ \partial r }{ \partial t } } = 
    { \frac{ 2^{ 1 / 4 } \lambda^{ 1 / 2 } }{ r^{ 1 / 4} } }, \quad \text{ as
    }  r \rightarrow \infty,
\label{eq.3.10}
\end{equation}

\noindent which has a solution

\begin{equation}
  r = \left( \frac{ 5 \lambda^{ 1 / 2}
    }{ 2^{ 7 / 4 } } \right)^{ 4 / 5 } t^{ 4 / 5 }.
\label{eq.3.11}
\end{equation}

\begin{figure}
  \begin{center}
    \includegraphics[scale=0.6]{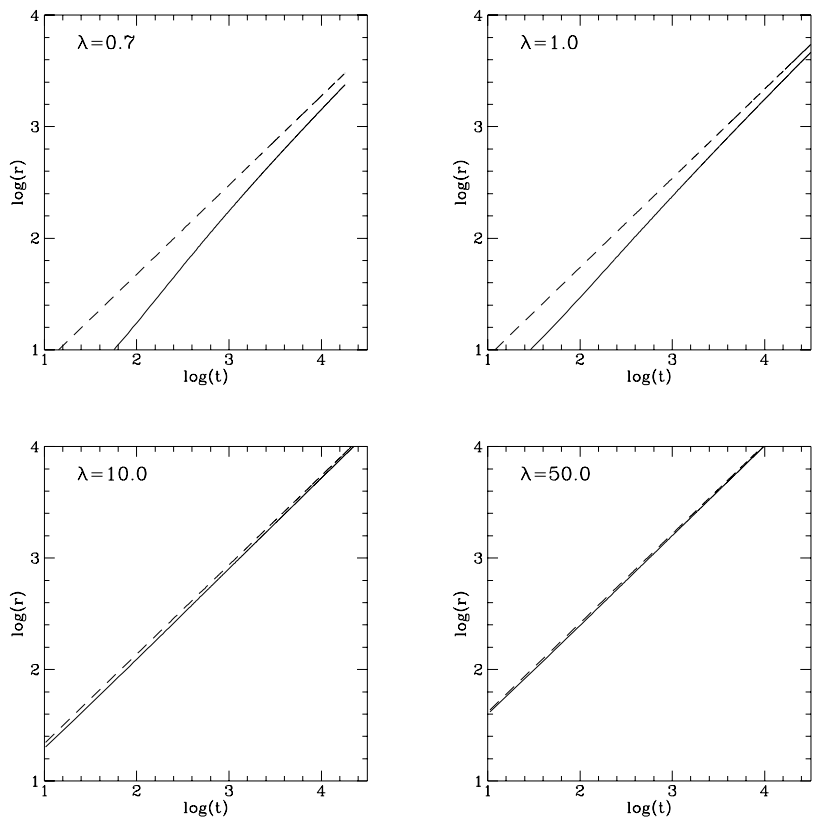}
  \end{center}
  \caption[Analytical approximation of the shock position for large
	   times]{ When \( \lambda > 1/2 \), the geometrical form of the
	   shock is spherical for sufficiently large times and its radial
	   coordinate grows as a power law (see Eq.~\eqref{eq.3.11}).
	   The curves on the plot show the shock position \( r \)
	   evaluated on the polar axis measured in units of the radius
	   of the disc \( r_\text{d} \).  Continuous lines show the
	   numerical calculation and dotted ones represent the analytical
	   approximation.  The time \( t \) is measured in units of \(
	   r_\text{d} / v_\text{k} \). }
\label{fig.3.2}
\end{figure}

  Figure~\ref{fig.3.2} shows for which times the approximation made with
Eq.~\eqref{eq.3.11} is valid on the polar axis of the shock.


\section{Astrophysical consequences}
\label{astrophysical-consequences}

  We have assumed that the shock produced by the stellar wind, as
well as the one produced by the accretion flow, occupy the same position
in space.  In this section we analyse under which circumstances this
approximation is correct.  As usual, we define the cooling length \(
l_\text{c} \) as the distance travelled by a particle from the shock
front until it reaches a point in which the temperature has a value \(
\sim 10^{4} \, \textrm{K} \) \citep{hartigan87}.

  \citet{hartigan87} estimated cooling lengths for different shock
velocities \( v_\text{sn} \) in the interval \(  20 \textrm{ km } 
\textrm{s}^{-1} \lesssim v_\text{sn} \lesssim 400 \textrm{ km }
\textrm{s}^{-1} \), and particle number density \( n \) in the
interval \( 100 \textrm{ cm }^{-3} \lesssim n \lesssim  1000 \textrm{
cm }^{-3} \).  From their results it follows that \citep{canto88}

\begin{gather}
  \left( \frac{ l_\text{c} }{ \textrm{AU} } \right) = A \left( \frac{
    v_\text{sn} }{ 100 \textrm{km} \textrm{s}^{-1} } 
    \right)^\beta  \left( \frac{ n }{ \textrm{cm}^{-3} } \right)^{ -1 },
						\label{eq.4.1} \\
  \intertext{where}
  A = 
    \begin{cases}
      6.6    & v_\text{sn} \lesssim 70  \textrm{km} \textrm{s}^{-1},	 \\ 
      43.33  & v_\text{sn} \gtrsim  70 \textrm{km} \textrm{s}^{-1}.
    \end{cases}
  						\label{eq.4.2} \\
  \intertext{and}
  \beta =
    \begin{cases}
      -5.5     &  v_\text{sn} \lesssim 70 \textrm{km} \textrm{s}^{-1}, \\
      5.7      &  v_\text{sn} \gtrsim  70 \textrm{km} \textrm{s}^{-1} 
    \end{cases}
   						\label{eq.4.3}
\end{gather}

\noindent  From now on, we will assume that Eq.~\eqref{eq.4.1} is valid
for any density and velocity ranges.

  The equations that describe the accretion flow
(Eqs.~\eqref{eq.1.13}-\eqref{eq.1.17}), the stellar wind flow
(Eq.~\eqref{eq.2.4}), the shock velocity (Eq.~\eqref{eq.3.4}) and the time
can be written in dimensional form as follows

\begin{gather}
  r = r_\text{d} \widetilde r, \quad 
    \rho = \rho_0 \widetilde \rho, \quad
    v_\text{i} = v_\text{k} \widetilde v_\text{i} \ \
    \text{(} i = r, \theta, \varphi \text{)}, \quad
    v_\text{s} = v_\text{k} \widetilde v_\text{s},
    							\notag \\
  v_\text{w} = v_\text{w} \widetilde v_\text{w}, \quad
    \rho_\text{w} = \rho_{\text{w}0} \widetilde \rho_\text{w},\quad
    t = \left( r_\text{d} / v_\text{k} \right) \widetilde t,
							\label{eq.4.4}
\end{gather}

\noindent where tilded quantities are referred to the dimensionless
variables considered in the previous sections.  We now give values of the
hydrodynamical quantities for the accretion flow and the stellar wind for
typical values of recently formed stars of low mass

\begin{gather}
  \left( \frac{ v_\text{k} }{ \textrm{km} \ \textrm{s}^{-1} } \right) 
    = 3 \left( \frac{ M }{ M_\odot } \right)^{ 1 / 2 } \left(
    \frac{ r_\text{d} }{ 100 \textrm{AU } } \right)^{ -1 / 2 } ,
					\label{eq.4.5} \\
  \begin{split}
    \left( \frac{ n_0 }{ \textrm{cm}^{-3} } \right) = 3.6 \times
      10^6 & \left( \frac{ \dot{M} }{  10^{ -6 }  \textrm{M}_\odot
      \text{yr}^{-1} } \right) \times \\
    & \times \left( \frac{ r_\text{d} }{ 100 \textrm{AU } } 
      \right)^{ -3 / 2 } \left( \frac{ M }{ \textrm{M}_\odot }
      \right)^{ -1 },
    \end{split}
  					\label{eq.4.6} \\
  \begin{split}
    \left( \frac{ n_{\text{w}0} }{ \textrm{cm}^{-3}  }
      \right) = 3 \times 10^4   & \left( \frac{ \dot M_\text{w} }{ 
      10^{ -7 } }{ \textrm{M}_\odot \text{yr}^{-1} } 
      \right) \times \\
    & \times \left( \frac{ r_\text{d} }{ 100 \textrm{ AU } } 
      \right)^{ -2 } \left( \frac{ v_\text{w} }{ 300 \textrm{km} \ 
      \textrm{s}^{-1} } \right)^{ -1 },
  \end{split}
    					\label{eq.4.7}
\end{gather}

\noindent where \( n_0 \) y \( n_{\text{w}0} \) are the particle number
densities of the accretion flow and the stellar wind respectively with the
assumption that the average mass per particle is \(\bar m \sim 1.3 m_H\).
As shown by Eq.~\eqref{eq.4.5}, typical values of the velocity \( v_\text{k}
\) are sufficiently small to be used in the calculation of the cooling
length of the accretion flow using Eq.~\eqref{eq.4.1}.  Thus, from now on,
we will only analyse the cooling lengths for the stellar wind.

  In order to simplify things further we calculate the cooling lengths for
the stellar wind leaving fixed the quantities \( M \), \( \dot{M} \), \(
r_d \), \( v_\text{w} \), with typical values as the ones showed in
Eqs.~\eqref{eq.4.5}--\eqref{eq.4.7} and we will only vary the quantity \(
\dot{M}_\text{w} \).  Table \ref{table4.1} shows the maximum variation of the 
ratio of the cooling length to the length between the shocks for different
polar angles in the steady case.  These results show that, for the steady
case the assumption of a thin post--shock layer is correct for the stellar
wind.


\begin{table}
  \begin{center}
    \begin{tabular}{|c|c|}
      \hline
      \hline
      \( \lambda \)  &  \( \left( l_\text{c} / r \right)_\text{max}
                           \times 10^3 \)   \\
      \hline
    
         0.04   &    0.54  \\
         
         0.08   &    1.08  \\
          
         0.12   &    1.65  \\
        
         0.16   &    2.25  \\  
        
         0.20   &    2.88  \\
        
         0.24   &    3.56  \\
        
         0.28   &    4.27  \\
        
         0.32   &    5.20  \\
       
         0.36   &    6.38  \\
       
         0.40   &    7.96  \\
       
         0.44   &   10.35  \\
        
         0.48   &   15.21  \\  
      \hline
    \end{tabular}
  \end{center}
  \caption[Cooling lengths]{Values of the cooling length in the
	 steady case for the shock produced by the stellar wind as a
	 function of the parameter \( \lambda \).  The quantity \( (l_c/
	 r)_\text{max}\) represents the maximum value of the ratio of
	 the cooling lengths \( l_c \) to the position of the shock \(
	 r \), for different polar angles.}
\label{table4.1} 
\end{table}

  Let us now analyse the cooling lengths of the stellar wind for the
case in which the shock evolves with time.  In the frame of reference
of the shock wave, the formulae used for the steady case are valid.
With the aid of Eq.~\eqref{eq.3.1},  in the frame of reference of the
central star, Eq.~\eqref{eq.4.1} takes the following form

\begin{equation}
  \left( \frac{ l_\text{c} }{ \textrm{AU} }\right) = A \left( \frac{
    \widetilde v_\text{wn} v_\text{w} - \widetilde v_\text{sn}
    v_\text{k} }{ 100  \textrm{km} \textrm{s}^{-1} }
    \right)^\beta   \left( \frac{ \widetilde \rho_\text{w} n_{\text{w}0}
    }{ \textrm{cm}^{-3} } \right)^{-1}.
\label{eq.4.8}
\end{equation}

\begin{figure}
  \begin{center}
    \includegraphics[scale=0.6]{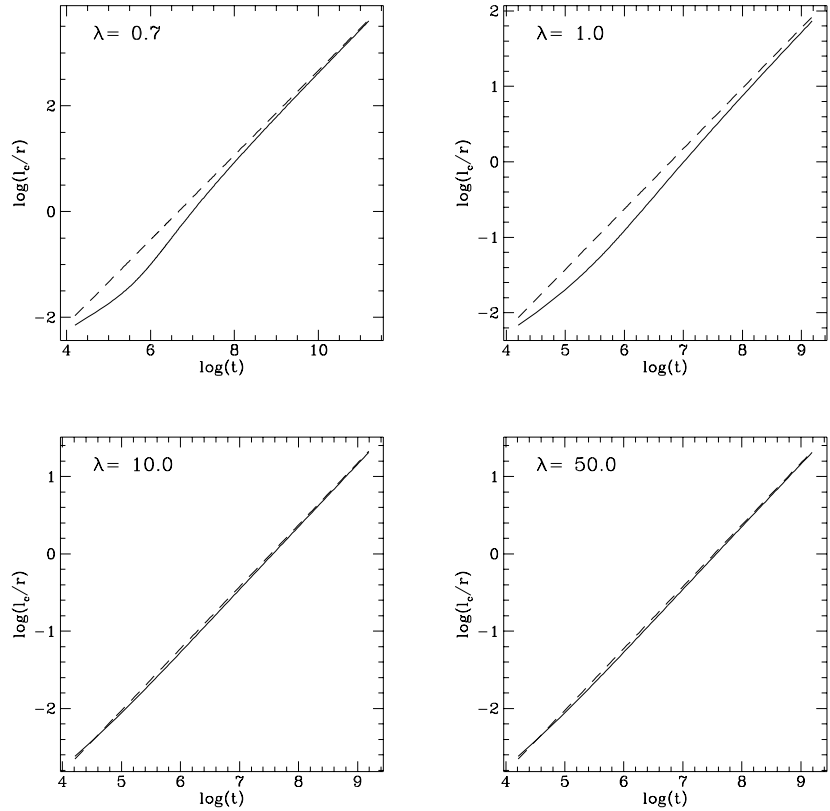}
  \end{center}
  \caption[Cooling lengths for	$\lambda >1/2$]{ When the parameter
	   \(\lambda >1/2\),  the thin layer approximation is no longer
	   valid for sufficiently large times.	The plot shows the
	   variation of the cooling length \((l_c)\) as a function of
	   time. The coordinate \(r\) is the distance from the star
	   to the shock. Dotted lines correspond to the asymptotic
	   approximation obtained in Eq.~\eqref{eq.4.9} and continuous
	   lines correspond to the numerical calculation.  The time \(t\)
	   is measured in years.}
\label{fig.4.1}
\end{figure}

  In this case, it follows that the ratio of the cooling length to
the radial distance to the star, when \(\lambda > 1/2\), is large.
To see this, let us make an asymptotic expansion of Eq.~\eqref{eq.4.8}
for \(r\gg 1\). With this, and using Eq.~\eqref{eq.3.11}, it is found that

\begin{equation}
  \left( \frac{ l_\text{c} }{ \textrm{ AU } } \right) \left( \frac{
    \widetilde r r_\text{d} }{ \textrm{ AU } } \right)^{ -1 } = A \left(
    \frac{ 5 \lambda^{ 1 / 2 } }{ 2^{ 7 / 4 } } \right)^{ 4 / 5 } \left(
    \frac{ v_\text{w} }{ 100 \textrm{km} \textrm{s}^{-1} } \right)^\beta
    \times \left( \frac{ n_{\text{w}0} }{ { \textrm{cm}^{-3} } } \right)^{
    -1 } \left(  \frac{ r_\text{d} }{ \textrm{ AU } } \right)^{-1}
    \widetilde t^{ 4 / 5 },
\label{eq.4.9}
\end{equation}

\noindent as \( t \rightarrow \infty\).  Figure~\ref{fig.4.1} shows at
which times the thin layer approximation ends its validity for certain
values of the parameter \(\lambda > 1/2\).  In the case where the shock
evolves with time, for \(\lambda \leq 1/2\), the cooling lengths and the
distance to the shock are always small until the steady configuration
is obtained, converging to the values shown on Table \ref{table4.1}.


  Let us now analyse the shape that a radio continuum observation will
show of the interaction described so far.  For this, we assume that the
shock emission is due to free--free processes and we use the brightness
temperature \( T_\text{B} \) given by

\begin{equation}
  T_B = T \left( 1 - e^{ -\tau_\nu } \right),
\label{eq.4.10}
\end{equation}

\noindent where \( T \)  is the temperature produced by the shocks.
The optical depth \(\tau_\nu\) in the continuum, at a radio frequency 
\(\nu\) of a shock wave with velocity \( v_\text{s} \) and a pre--shock
density \(n\) is given by \citep{curiel89}

\begin{equation}
  \begin{split}
    \tau_\nu = 9.83 \times 10^{-7}
      & \left( \frac{ n }{ \textrm{cm}^{-3} } \right) \left( \frac{
        v_s }{  300  \textrm{km} \textrm{s}^{-1} } \right)^{ 1.68 }
        \times \\
      & \times \left( \frac{ T }{ 10^{4} \textrm{K} } \right)^{-0.55}
        \left( \frac{ \nu }{ 5 \textrm{ GHz } } \right)^{-2.1}.
  \end{split}
\label{eq.4.11}
\end{equation}

  For our model, it follows that \( \tau_\nu \ll 1 \) and so,
Eq.~\eqref{eq.4.10} can be written as

\begin{equation}
  {T_B} =  T \tau_\nu, \quad \text{ as } \tau_\nu \rightarrow 0.
\label{eq.4.12}
\end{equation}

\begin{figure}
  \begin{center}
    \includegraphics[scale=0.6]{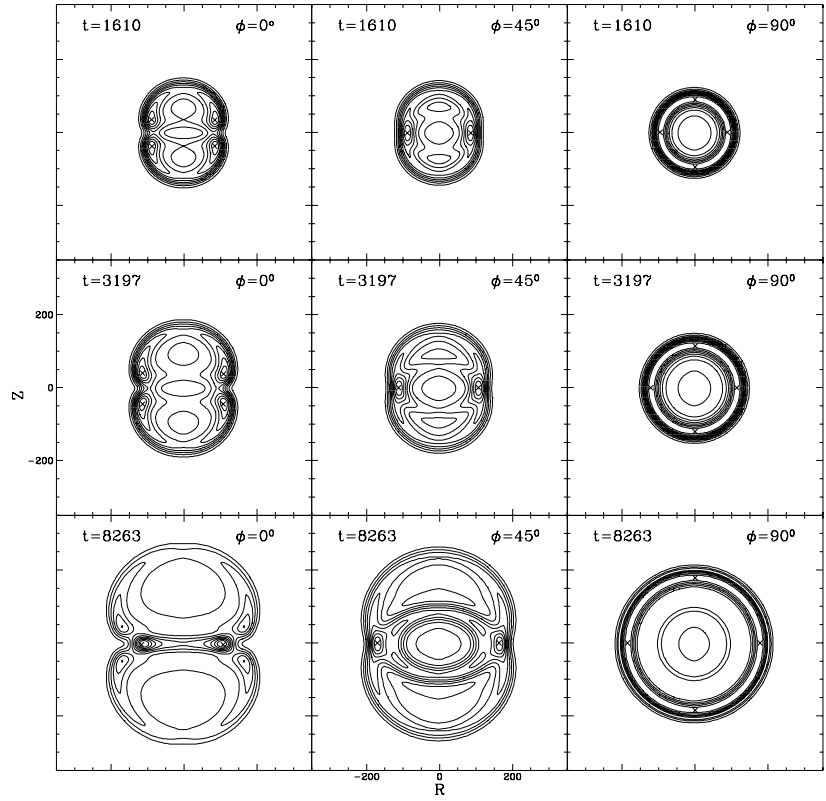}
  \end{center}
  \caption[Shock emission at different times]{ Emission isocontours
           normalised to the maximum (shown with crosses).  The isocontours
	   take values \(0.1,\, 0.2,\, \ldots 1.0 \) with
           \(\dot M_w =10^{-7} \textrm{M}_\odot \textrm{yr}^{-1}\)
           (\(\lambda =10\)) and a resolution \(\theta_B =0''.2\) at a
	   distance of \(150 \textrm{pc} \).   The angle \(\phi\) is
	   formed between the equatorial plane and the observer.
           The time \(t\) is measured in years.  The cylindrical radius 
	   \( R \) and the coordinate \( Z \) are measured in astronomical
	   units.}
\label{fig.4.2}
\end{figure}

\begin{figure}
  \begin{center}
    \includegraphics[scale=0.6]{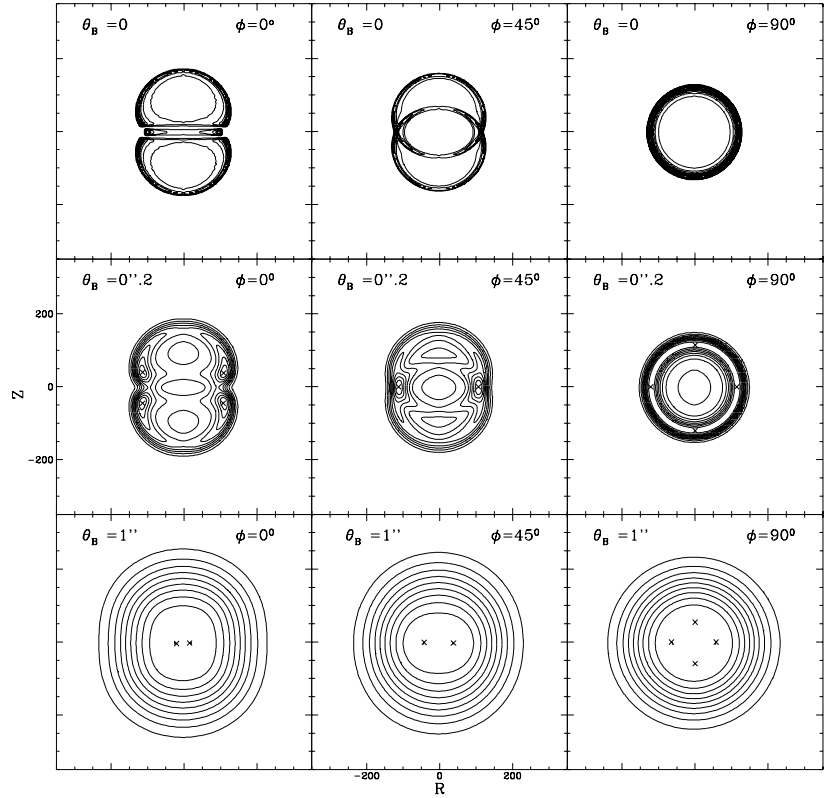}
  \end{center}
  \caption[Shock emission for different resolutions]{ Emission isocontour
	   variations due to different resolutions, normalised to the
	   maximum (shown with crosses).  The isocontours take values \(
	   0.1,\, 0.2,\, \ldots 1.0 \) with \(\dot M_\text{w} =10^{-7}
	   \textrm{M}_\odot \textrm{yr}^{-1} \) \((\lambda =10)\) and
	   \(t = 3197\) yrs.  The angle \(\phi\) is measured is the
	   angle formed by the equatorial plane and the line of sight.
	   The angle \(\theta_\text{B}\) represents the resolution.
	   The cylindrical radius \( R \) and the coordinate \( Z \)
	   are measured in astronomical units. }
\label{fig.4.3}
\end{figure}

  Figures~\ref{fig.4.2}-\ref{fig.4.3} show different projections of the
brightness temperature isocontours on the plane of the sky for typical
values of \(\dot{M}_w =10^{-7} \textrm{M}_\odot \textrm{yr}^{-1} \) 
(\(\lambda = 10\)).

  Let us now calculate the emission flux produced by the stellar wind shock
for the value used in Figures~\ref{fig.4.2}-\ref{fig.4.3} in which 
\(\dot M_\text{w} =10^{-7} \textrm{M}_\odot \textrm{yr}^{-1}\) \((\lambda
=10)\).  For this case, the emission flux received on the plane of the sky
is given by \citep{rodriguez90}

\begin{equation}
  S_\nu = \frac{ 2 k T \nu^2 }{ \mathrm{c}^2 } \int{ \mathrm{d} \Omega 
    \tau_{\nu o} },
\label{eq.4.13}
\end{equation}

\noindent where the integral is extended over the solid angle \(
\Omega \) occupied by the shock surface on the plane of the sky and
\( \mathrm{c} \) is the speed of light.  The optical depth \(\tau_{\nu 0}\)
is observed through the plane of the sky.  The quantity \(\tau_{\nu 0}
d\Omega = \tau_\nu \mathrm{d}a / D^2\), where \( \mathrm{d}a \) is the
area element and \(\tau_\nu\) is the optical depth, both normal to the
shock surface. With this and because \( D \) is the distance from the
observer to the shock surface, Eq.~\eqref{eq.4.13} can be rewritten as

\begin{equation}
  S_\nu = \frac{ 2 K T \nu^2 }{ D^2 \mathrm{c}^2} \int{ \mathrm{d}a \tau_\nu }.
\label{eq.4.14}
\end{equation}

  Using Eq.~\eqref{eq.2.7} in spherical coordinates, Eq.~\eqref{eq.4.14}
takes the form

\begin{equation}
  S_\nu =   \frac{ 2 K T \nu^2 }{ D^2 \mathrm{c}^2 } \oint{  \frac{
    \mathrm{d} \theta \mathrm{d} \varphi \tau_\nu r \sin \theta }{ \sqrt{
    r^2 + { \left( \partial r / \partial \theta \right)^2 } } } \left\{
    r^2 - \left( \partial r / \partial \theta \right)^2 \right\} }.
\label{eq.4.15} 
\end{equation}

\noindent  In our case, Eq.~\eqref{eq.4.15} is

\begin{gather}
  \left( \frac{ S_\nu }{ \textrm{mJy} } \right) = 0.406 
    \left( \frac{ T }{ 10^{4} \textrm{K} } \right)
    \left( \frac{ \nu }{ 5 \textrm{ GHz } } \right)^2 
    \left( \frac{ D }{  150 \textrm{ pc } } \right)^{-2} \Phi,
					\label{eq.4.16} \\
  \intertext{where}
  \Phi \equiv  \int^{ \pi / 2 }_0{  \frac{ \mathrm{d} \theta \tau_\nu
    r \sin \theta }{ \sqrt{ r^2 + \left( \partial r  / \partial \theta
    \right)^2 } } \left\{ r^2 - \left( \partial r / \partial \theta 
    \right)^2 \right\} },
					\label{eq.4.17}
\end{gather}

\noindent and \( r \) is measured in astronomical units.

  When \(\lambda > 1/2\), because the derivative of the shock with respect
to the polar angle vanishes for sufficiently large times, the emission flow
tends to a constant value given by

\begin{equation}
  \begin{split}
    \left( \frac{ S_\nu }{ \textrm{mJy} } \right) =   3.37  & \left( \frac{ \dot{M}_\text{w} }{ 10^{-7} }{
      \textrm{M}_\odot \text{yr}^{ -1 } } \right) \left( \frac{
      v_\text{w} }{ 300 \textrm{km} \textrm{s}^{-1}  } 
      \right)^{ 1.68 } \times  \\
    & \times \left( \frac{ T }{ 10^{4} }{ \textrm{K} } 
      \right)^2 \left( \frac{ \nu }{ 5 \textrm{GHz} } \right)^{ -0.1
      } \left( \frac{ D }{ 150 \textrm{ pc } } \right)^{-2}, \quad \text{
      as } t \rightarrow \infty.
  \end{split}
\label{eq.4.18}
\end{equation}

\begin{figure}
  \begin{center}
    \includegraphics[scale=0.6]{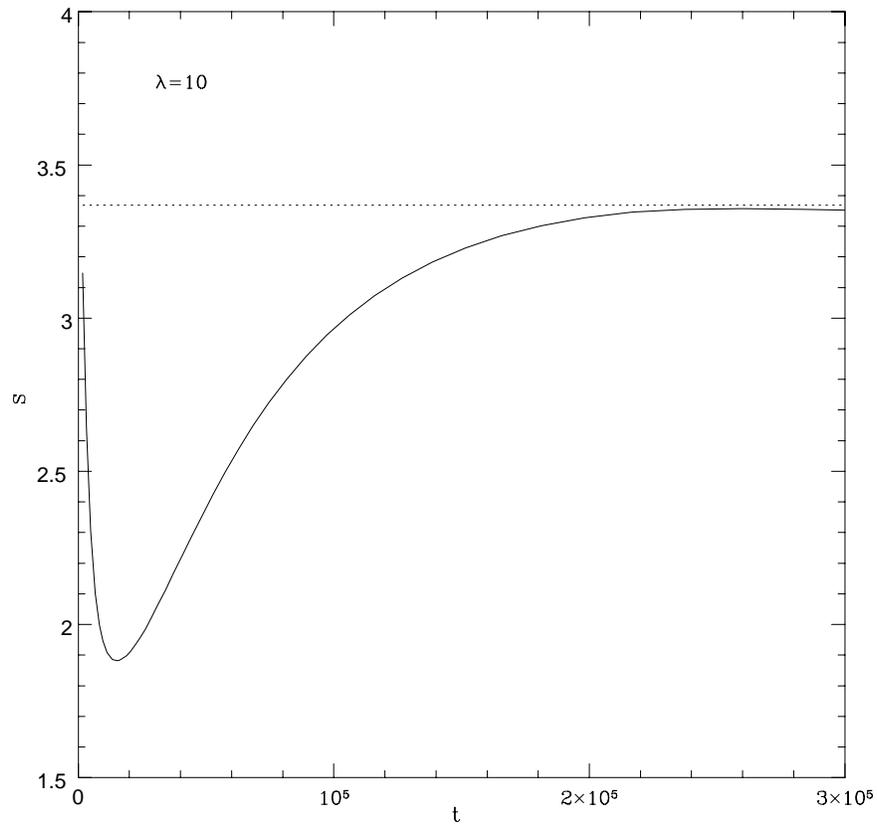}
  \end{center}
  \caption[Shock emission produced by the stellar wind]{ When the
           dimensionless parameter \(\lambda > 1/2\), the emission flux 
	   \( S \) tends to a constant value for \( t \gg 1 \).  The
	   continuous line represents the numerical calculation and the
	   dotted line the asymptotic analytical value.
           The time \( t \) is measured in years and the flux \( S \) 
	   in \(mJy\). }
\label{fig.4.4}
\end{figure}

  Figure~\ref{fig.4.4} shows the emission flux as a function of time
for \(\dot M_\text{w} =10^{-7} \textrm{M}_\odot \textrm{yr}^{-1}\)
\((\lambda =10)\).


\section{Conclusions}

  We have developed a simple prescription to describe the hydrodynamical
interaction between a rotating accretion flow onto a star with a
spherically symmetric wind.  For simplicity we considered a purely
hydrodynamical flow, with no magnetic fields.  The evolution of the shock
wave produced by the interaction of the wind with the accretion flow is
thus described by balancing the wind's force \( \sim \dot{M}_\text{w}
v_\text{w} \) with the accretion flow's force  \( \sim \dot{M} v_\text{k}
\), where \( v_\text{k} \) is the keplerian velocity that a particle
has at a characteristic distance of \( r_\text{d} \), the radius of the
disc.  That allowed us to define a dimensionless parameter \( \lambda
\) given by Eq.~\eqref{eq.2.5}.  For very large values of \( \lambda
\), the surface of interaction expands to infinity and the thin layer
approximation breaks down.  For small values of \( \lambda \), in fact
for \( \lambda < 1 / 2 \), it is possible to reach a steady configuration.

  We would like to point out that a full analysis of the interaction
should take into account the gravitational pull that the central star
makes on the gas and the equilibrium equation must also include a
centrifugal term.  If magnetic fields are to be included on the whole
description of the problem as well, then magnetic pressure will also play
an important role on the structure of the cavities.  We will make 
a more precise analysis considering all these physical processes in a
future article.

\bibliographystyle{phdthesis}
\bibliography{hydro}

\renewcommand{\theequation}{A\arabic{equation}}
\setcounter{equation}{0}
\section*{Appendix}
\label{appendix}

 Eq.\eqref{eq.1.18} is a reduced Cardan third order algebraic equation, i.e.

\begin{equation}
  x^3 \pm 3 \alpha x + 2 \beta = 0,
\label{eq.A.1}
\end{equation}

\noindent with

\begin{equation}
  \alpha = | \frac{ r - 1 }{ 3 }  |, \qquad \beta = - \frac{ r \cos \theta
    }{ 2 }, \quad \text{and} \quad  x = \cos \theta_0.
\label{eq.A.2}
\end{equation}

\noindent The solutions of Eq.~\eqref{eq.A.2} are given by
\citep{namias84}

\begin{equation}
  x = 
    \begin{cases}
      2 \alpha^{1/2} \sinh \left\{ \frac{ 1 }{ 3 } \textrm{arcsinh} \left( -
        \beta / \alpha^{3/2} \right) \right\}, \\
       \qquad \qquad \qquad \qquad \text{for a } + \text{ sign}, \\
      2 \alpha^{ 1 / 2 } \cosh \left\{ \frac{ 1 }{ 3 } \text{arccosh} \left(
	   - \beta / \alpha^{ 3 / 2 } \right) \right\}, \\
       \qquad \qquad \qquad \qquad \text{for a } - \text{ sign and } 
	   \beta^2 - \alpha^3 >0, \\
      2 \alpha^{ 1 / 2 } \cos \left\{ \frac{ 1 }{ 3 } \arccos \left(
         - \beta / \alpha^{ 3 / 2 }  \right) \right\}, \\
	\qquad \qquad \qquad \qquad \text{for a } - \text{ sign \& } 
	 \beta^2 - \alpha^3 < 0.
    \end{cases}
\label{eq.A.3}
\end{equation}
  
  For each case in Eq.~\eqref{eq.A.3} there exist three different solutions
due to the fact that

\begin{gather}
  \arccos z  = \pm \arccos_\text{p} z + 2 k \pi, 	
  						\notag \\
  \text{arcsinh} z = (-)^k \text{arcsinh}_\text{p} z + 2 k \pi i,
    						\notag \\
  \text{arccosh} z = \pm \text{arccosh}_\text{p} z + 2 k \pi i, 
    						\label{eq.A.5}
\end{gather}

\noindent with \( k = 0,\; \pm 1,\; \pm 2, \ldots \) The subindex p
refers to the main Riemann sheet.

  For our particular case, we are interested in real solutions, so for the
second and third cases in Eq.~\eqref{eq.A.5} we set \( k = 0 \).  For the
first case, due to the symmetry of the problem, we consider only values of
the polar angle such that \( 0 \leq \theta  \leq \pi / 2 \).  Let us define 
\( \mu \equiv - \beta / \alpha^{3/2} \) and analyse the behaviour of the
function \( \zeta = \cos \left\{ ( \arccos \mu ) / 3 + 2 k \pi \right\}
\) with \( k = 0,\ \pm 1 \).  When \( \mu > 0 \) we expect \( \zeta \geq 0
\).  Since \( \zeta( \mu \geq 0,\, k = 0 ) > 0 \) then the required
solution occurs when \( k = 0 \).  In other words, the required full 
solution is given by Eq.~\eqref{eq.A.3} with \( k = 0 \).

\eject

\eject

\eject

\eject

\eject

\eject

\eject

\eject

\eject

\eject

\eject

\eject

\eject

\end{document}